\documentclass{aastex}
\usepackage{amsmath, natbib, ulem}
\usepackage{graphicx}
\usepackage[usenames,dvips]{color}

\slugcomment{To be submitted to the Astrophysical Journal}

\begin{document}

\title{Interacting Binaries with Eccentric Orbits II.  Secular 
Orbital Evolution Due To Non-Conservative Mass Transfer} 

\author{J. F. Sepinsky$^{1,2}$, B. Willems $^2$, V. Kalogera$^2$, F. 
A. Rasio$^2$}
\affil{$^1$ The University of Scranton, Department of Physics and 
Electrical Engineerings, Scranton, PA 18510}
\affil{$^2$ Department of Physics and Astronomy, Northwestern 
University,
2145 Sheridan Road, Evanston, IL 60208}
\slugcomment{j-sepinsky, b-willems, vicky, and rasio@northwestern.edu}
\shortauthors{Sepinsky et al.}
\shorttitle{Interacting Binaries with Eccentric Orbits}

\begin{abstract}

We investigate the secular evolution of the orbital semi-major axis and
eccentricity due to mass transfer in eccentric binaries, allowing for
both mass and angular momentum loss from the system. Adopting a delta
function mass transfer rate at the periastron of the binary orbit, we
find that, depending on the initial binary properties at the onset of
mass transfer, the orbital semi-major axis and eccentricity can either
increase or decrease at a rate linearly proportional to the magnitude of
the mass transfer rate at periastron. The range of initial binary mass
ratios and eccentricities that leads to increasing orbital semi-major
axes and eccentricities broadens with increasing degrees of mass loss
from the system and narrows with increasing orbital angular momentum
loss from the binary. Comparison with tidal evolution timescales shows
that the usual assumption of rapid circularization at the onset of mass
transfer in eccentric binaries is not justified, irrespective of the
degree of systemic mass and angular momentum loss. This work extends our
previous results for conservative mass transfer in eccentric binaries
and can be incorporated into binary evolution and population synthesis
codes to model non-conservative mass transfer in eccentric binaries. 

\end{abstract}


\section{Introduction}
\label{sec-intro}

Many binary systems pass through at least one mass-transfer phase during
the course of their evolution.  These mass transfer episodes not only
affect the internal evolution of the stellar components, but also impact
the binary properties. In particular, changes in the mass ratio and
transfer of linear and angular momenta between the stars cause changes
in the orbital elements which affect the evolution of the binary and can
cause a feedback on the mass transfer process. 

In binaries with eccentric orbits, the stars are closest to each other
at the periastron of their relative orbit, so that any mass transfer is
expected to take place first during periastron passage. The standard
assumption in current binary evolution and population synthesis codes to
deal with such mass transfer phases is that the orbit circularizes
instantaneously at the onset of mass transfer. This assumption is in
contrast with recent theoretical findings \citep{2007ApJ...667.1170S} as
well as observations of semi-detached binaries with non-zero orbital
eccentricities \citep{1999AJ....117..587P, 2005A&AT...24..151R}.

\citet[][hereafter Paper~I]{2007ApJ...667.1170S} studied the orbital evolution
due to mass transfer in eccentric binaries by deriving a set of
perturbed equations of motion for the binary components, as outlined
initially by \citet{1969Ap&SS...3..330H}. The authors found that, under the
assumption of conservation of total system mass and orbital angular
momentum, the orbital semi-major axis and eccentricity can increase as
well as decrease, depending on the initial orbital elements, binary
component masses, and donor rotation rate. Furthermore, the orbital
evolution timescales can be short enough to compete with or enhance any
tidally driven orbital evolution.

In this paper, we extend the analysis presented in Paper~I to account
for mass and angular momentum loss from the binary. In
\S\,\ref{sec-basic} and \ref{sec-sec}, we briefly recall the relevant
ingredients for the study of mass transfer in eccentric binaries derived
in Paper~I, and update the formalism to account for systemic mass and
angular momentum loss. In \S\ref{orbevtim}, we present timescales for
the evolution of the orbital semi-major axis and eccentricity due to
mass transfer for different degrees of mass and angular momentum loss
and we compare the timescales to the timescales of orbital evolution due
to tidal dissipation. 
The final section is devoted  to concluding remarks.

\section{Basic Assumptions}
\label{sec-basic}

We consider a close binary consisting of two stars with masses $M_1$ and
$M_2$ in an eccentric orbit with period $P_{\rm orb}$, semi-major axis
$a$, and eccentricity $e$. The stars are assumed to rotate uniformly
with angular velocities $\vec{\Omega}_1$ and $\vec{\Omega}_2$ around an
axis perpendicular to the orbital plane and in the same sense as the
orbital motion. Since, in eccentric binaries, the magnitude of the
orbital angular velocity $\vec{\Omega}_{\rm orb}$ is a periodic function
of time, the stellar rotation rates cannot be synchronized with the
orbital motion at all orbital phases. 

At some time $t$, one of the stars is assumed to fill its Roche lobe
initiating mass transfer to its companion through the inner Lagrangian
point $L_1$. We assume this point to lie on the line connecting the mass
centers of the stars, even though non-synchronous rotation may cause it
to oscillate in the direction perpendicular to the orbital plane, as
shown in the appendix of \citet{MW83}. Since the donor's equatorial
plane coincides with the plane of the orbit, the transferred mass can
furthermore be assumed to remain in the orbital plane at all times. In
what follows, we refer to the Roche lobe filling star as star~1 and to
the companion star as star~2.

In Paper~I, we assumed that all mass transfered by the donor was
accreted by the companion, and that any orbital angular momentum
transported by the transferred matter was immediately returned to the
orbit (presumably through an accretion disk). 
Here, we relax those assumptions and assume some fraction $\beta$ of the
transferred mass to be lost from the system: 
\begin{equation}
\dot{M}_T = \beta\, \dot{M}_1,
\end{equation}
where $M_T = M_1+M_2$ is the total system mass. Consequently, the amount of 
matter accreted by the companion is
\begin{equation}
\dot{M}_2 = -\gamma\, \dot{M}_1
\label{gammadef}
\end{equation}
where $\gamma=1-\beta$.  The angular momentum carried away by the mass
lost from the system is parameterized in terms of the specific angular 
momentum of the orbit as
\begin{equation}
\dot{J}_{\rm orb} = \mu\, \frac{J_{\rm orb}}{M_T}\, \dot{M}_T. 
\label{mudef}
\end{equation}
In the particular case where the matter lost from the system carries the 
specific orbital angular momentum of the accretor, $\mu = M_1/M_2$ 
\citep[e.g.,][]{Kolb2001}. For our purpose, we  
assume that no other sources of angular momentum loss besides mass loss 
are operating on the system. Any other sinks of 
orbital angular momentum such as tidal interactions, magnetic braking, and 
gravitational radiation can, at the lowest order of approximation, be 
added to Eq.~(\ref{mudef}) and Eqs.~(\ref{eq-deltaa})--(\ref{eq-deltae}) 
derived in the next section to obtain the total rate of change of the 
orbital angular momentum and of the orbital elements.

With these assumptions, the equations governing the motion of the two
stars around their common center of mass can be written in the form of a
perturbed two-body problem as
\begin{equation}
\frac{d^2\vec{r}}{dt^2} = -\frac{G \left( M_1+M_2 \right)}
  {\left| \vec{r} \right|^3}\,
  \vec{r} + S\, \vec{\hat{x}} + T\, \vec{\hat{y}} + W\, \vec{\hat{z}},
  \label{pertmot}
\end{equation}
where $G$ is the Newtonian constant of gravitation, $\vec{r}$ is the 
position vector of the accretor with respect to the donor, 
$\vec{\hat{x}}$ is a unit vector in the direction of $\vec{r}$,
$\vec{\hat{y}}$ is a unit vector in the orbital plane perpendicular to
$\vec{r}$ in the direction of the orbital motion, and $\vec{\hat{z}}$ is 
a
unit vector perpendicular to the orbital plane parallel to and in the
same direction as $\vec{\Omega}_{\rm orb}$. The functions $S$, $T$,
and $W$ are the components of the perturbing force arising from the
mass transfer between the binary components. In Paper I, we have shown 
that they are equal to
\begin{eqnarray}
S &=& \frac{f_{2,x}}{M_2} - \frac{f_{1,x}}{M_1}  
+ \frac{\dot{M}_2}{M_2} \left( v_{\delta M_2,x} - |\vec{\Omega}_{\rm orb}|
|\vec{r}_{A_2}| \sin{\phi} \right) \nonumber \\
&-& \frac{\dot{M}_1}{M_1} v_{\delta M_1,x} 
+ \frac{\ddot{M}_2}{M_2} |\vec{r}_{A_2}| \cos{\phi}
- \frac{\ddot{M}_1}{M_1} |\vec{r}_{A_1}|, \label{eq-S} \\
T &=& \frac{f_{2,y}}{M_2} - \frac{f_{1,y}}{M_1}  
+ \frac{\dot{M}_2}{M_2} \left( v_{\delta M_2,y} 
+ |\vec{\Omega}_{\rm orb}| |\vec{r}_{A_2}| \cos{\phi} \right) \nonumber \\ 
&-&\frac{\dot{M}_1}{M_1} \left( v_{\delta M_1,y}
+ |\vec{\Omega}_{\rm orb}||\vec{r}_{A_1}| \right) +
\frac{\ddot{M}_2}{M_2} |\vec{r}_{A_2}| \sin{\phi}, \label{eq-T} \\
W &=& \frac{f_{2,z}}{M_2} - \frac{f_{1,z}}{M_1}, \label{eq-W}
\end{eqnarray}
where $A_1$ denotes the point on the donor's surface from which mass is
lost (the $L_1$ point), $A_2$ denotes the point on the accretor's
surface at which mass is accreted, $\vec{r}_{A_1}$ and $\vec{r}_{A_2}$
are the position vectors of $A_1$ and $A_2$ with respect to the donor's
and the accretor's center of mass, respectively, and the subscripts $x$,
$y$, and $z$ denote vector components in the $\vec{\hat{x}}$, 
$\vec{\hat{y}}$, and $\vec{\hat{z}}$ directions. Moreover, $\vec{f}_1$ 
is the gravitational force exerted by particles in the mass transfer 
stream on the donor star, $\vec{f}_2$ the gravitational force exerted by 
particles in the mass transfer stream on the accretor, $\vec{v}_{\delta 
M_1}$ the velocity of the matter ejected at $L_1$ with respect to the 
mass center of the donor star, $\vec{v}_{\delta M_2}$ the velocity of 
the accreted matter at $A_2$ with respect to the mass center of the 
accretor, and $\phi$ the angle between $\vec{\hat{x}}$ and the vector 
from the center of mass of the accretor to $A_2$.  More details, as well 
as a diagram relating the above vectors, can be found in Paper~I.

\section{Orbital Evolution Equations}
\label{sec-sec}

\subsection{Secular Variation of the Orbital Elements}

The perturbing force with components $S$, $T$, and $W$ in the equations
governing the motion of the two stars around their common center of mass
causes changes in the orbital semi-major axis $a$ and eccentricity $e$
at rates given by \citep[e.g.,][]{S60, BC61, D62, F70}
\begin{equation}
\frac{da}{dt} = \frac{2}{n(1-e^2)^{1/2}} [ S e \sin{\nu} + T ( 1
  + e \cos{\nu} )],
\label{eq-dadt}
\end{equation}
\begin{eqnarray}
\lefteqn{\frac{de}{dt} = \frac{(1-e^2)^{1/2}}{na}} 
  \nonumber \\
 & & \times \left\{ S \sin{\nu} +
  T \left[ \frac{2\cos{\nu} +e \left(1+\cos^2{\nu}
  \right)}{1+e\cos{\nu}} \right] \right\},
\label{eq-dedt}
\end{eqnarray}
where $n=2\pi/P_{\rm orb}$ is the mean motion, and $\nu$ the true
anomaly. The long-term secular evolution of the orbital
elements is obtained by averaging these equations over one orbital 
period:
\begin{equation}
\left< {\frac{da}{dt}} \right>_{\rm sec} \equiv 
  \frac{1}{P_{\rm orb}} \int_{-P_{\rm orb}/2}^{P_{\rm orb}/2}
  {\frac{da}{dt}}\, dt,
\label{eq-dadtsec}
\end{equation}
\begin{equation}
\left< {\frac{de}{dt}} \right>_{\rm sec} \equiv 
  \frac{1}{P_{\rm orb}} \int_{-P_{\rm orb}/2}^{P_{\rm orb}/2}
  {\frac{de}{dt}}\, dt.
\label{eq-dedtsec}
\end{equation}
As in Paper~I, the integrals are most conveniently computed in
terms of the true anomaly $\nu$. 

\subsection{Orbital Angular Momentum Loss}
\label{sec-angmom}

The perturbing functions $S$ and $T$ depend on the properties of the
mass transfer stream. Calculation of the orbital semi-major axis and
eccentricity evolution therefore, in principle, requires the calculation
of the trajectories of the particles in the stream. In Paper~I we
bypassed such a calculation by assuming conservation of total system
mass and orbital angular momentum. Here, we relax this assumption and
generalize the formalism presented in Paper~I by parameterizing systemic
mass and angular momentum loss by means of Eqs.~(\ref{gammadef}) and
(\ref{mudef}). 

As shown in Paper~I, the rate of change of the orbital angular momentum
is related to the perturbing function $T$ by
\begin{equation}
\frac{\dot{J}_{\rm orb}}{J_{\rm orb}} = \frac{\dot{M_1}}{M_1} +
\frac{\dot{M_2}}{M_2}  
- \frac{1}{2}\frac{\dot{M_T}}{M_T}
+ \frac{\left( 1-e^2 \right)^{1/2}}{n a (1+e\, \cos \nu)}\, T.
\label{eq-Jdot}
\end{equation}
If the only sink of orbital angular momentum is mass loss from the system, 
elimination of $\dot{J}_{\rm orb}$ from this equation and  
Eq.~(\ref{mudef}) yields
\begin{equation}
\left(\mu + \frac{1}{2} \right) \frac{\dot{M}_T}{M_T} - \frac{\dot{M_1}}{M_1} 
- \frac{\dot{M_2}}{M_2}  = 
\frac{\left( 1-e^2 \right)^{1/2}}{n a (1+e\, \cos \nu)}\, T.
\label{jorbmu}
\end{equation}

\section{Orbital Evolution Timescales}
\label{orbevtim}

Since in binaries with eccentric orbits, mass transfer is expected to
occur first at the periastron of the binary orbit, we approximate the
mass transfer rate by a Dirac delta function as
\begin{equation}
\dot{M}_1 = \dot{M}_0\, \delta \left( \nu \right),
\label{eq-del}
\end{equation}
where $\dot{M}_0 < 0 $, $|\dot{M}_|$ is the instantaneous mass 
loss rate of star~1, and $\delta(\nu)$ the Dirac delta function.

To calculate the rates of secular change of the orbital semi-major axis
and eccentricity, we neglect the gravitational force exerted by the
particles in the mass-transfer stream on the binary components and set
\begin{eqnarray}
f_{1,x} &=& f_{2,x} = 0, \label{eq-f12x} \\
f_{1,y} &=& f_{2,y} = 0. \label{eq-f12y}
\end{eqnarray}
Since the particles are confined to the orbital plane, the $z$-component 
of the perturbing forces are zero as well.  A calculation of the 
actual gravitational effect of the particles in the mass transfer stream 
is beyond the scope of this paper and will be left for future studies.

With these assumptions, elimination of the perturbing function $T$ between 
Eqs.~(\ref{eq-T}) and~(\ref{jorbmu}), substituting Eqs. (\ref{gammadef}) 
and (\ref{mudef}), and 
averaging over one 
orbital period yields
\begin{eqnarray}
\lefteqn{\gamma qv_{\delta M_2,y} + v_{\delta M_1,y} = 
- |\vec{\Omega}_{{\rm orb},P}||\vec{r}_{A_1,P}|} \nonumber \\
&-& na\left(\frac{1+e}{1-e}\right)^{1/2} \left[ \gamma q - 1 + (1-\gamma)\left(\mu + 
\frac{1}{2}\right)\left(\frac{q}{1+q}\right)\right]\nonumber \\
&-& \gamma q|\vec{\Omega}_{{\rm orb}, P}||\vec{r}_{A_2}|\cos{\phi_P} 
\left[ 1- \left( \frac{d\phi}{d\nu}\right)_P \right], 
\label{eq-vr}
\end{eqnarray}
where the subscript $P$ indicates quantities evaluated at periastron,
and $q=M_1/M_2$ is the binary mass ratio. In the limiting case of
conservative mass transfer ($\gamma=1$), Eq.~(\ref{eq-vr}) reduces to
Eq.~(35) in Paper~I.

Substituting Eqs.~(\ref{eq-f12x})--(\ref{eq-vr}) into Eqs.~(\ref{eq-S})
and (\ref{eq-T}) yields expressions for the $S$ and $T$ components of
the perturbing force acting on the binary due to the mass transfer
between the binary components.With these expressions,
Equations~(\ref{eq-dadt})--(\ref{eq-dedtsec}) yield the following rates
of secular change of the orbital semi-major axis and eccentricity due to
mass transfer at the periastron of eccentric binaries:
\begin{eqnarray}
\left< \frac{da}{dt} \right>_{\rm sec} = &&\frac{a}{\pi} 
\frac{\dot{M}_0}{M_1} \frac{1}{(1-e^2)^{1/2}} \nonumber \\
   &\times&\left[ e \frac{|\vec{r}_{A_1,P}|}{a} + 
\gamma qe\frac{|\vec{r}_{A_2}|}{a}\cos{\Phi_P} \right.\nonumber\\ &+& 
(\gamma q-1)(1-e^2)\nonumber \\ &+& \left.
(1-\gamma)(\mu+\frac{1}{2})(1-e^2)\frac{q}{1+q}   
\right],
\label{eq-deltaa}
\end{eqnarray}
\begin{eqnarray}
\left< \frac{de}{dt} \right>_{\rm sec} = &&\frac{(1-e^2)^{1/2}}{2\pi}
  \frac{\dot{M}_0}{M_1} \nonumber \\
 &\times& \left[ \gamma q 
\frac{|\vec{r}_{A_2}|}{a}\cos{\Phi_P} + \frac{|\vec{r}_{A_1,P}|}{a} 
\right. \nonumber \\ 
&+& 2(\gamma q - 1)(1-e) \nonumber \\
&+& \left. 2(1-\gamma)(\mu+\frac{1}{2}) (1-e)\frac{q}{1+q} \right].
\label{eq-deltae}
\end{eqnarray}
The rates of orbital evolution are thus linearly proportional to the
magnitude $\dot{M}_0$ of the mass loss rate at periastron and
dependent on the orbital semi-major axis $a$, the orbital eccentricity
$e$, the donor mass $M_1$, and the binary mass ratio $q$.  They also
depend on the donor's rotational angular velocity $\Omega_1$ through the
position vector $\vec{r}_{A_1}$ of the inner Lagrangian point $L_1$
\citep[e.g., ][]{2007ApJ...660.1624S}. Since $|\vec{r}_{A_1, P}| \propto 
a$, the timescales explicitly depend on the orbital semi-major axis $a$ 
only through the ratio $|\vec{r}_{A_2}|/a$ of the accretor's equatorial
radius to the orbital semi-major axis. In the limiting case of 
conservative mass transfer ($\gamma=1$), Eqs.~(\ref{eq-deltaa})
and~(\ref{eq-deltae}) reduce to Eqs.~(39) and~(40) in Paper~I. 

\begin{figure*}
\begin{center}
\resizebox{8.5cm}{!}{\includegraphics{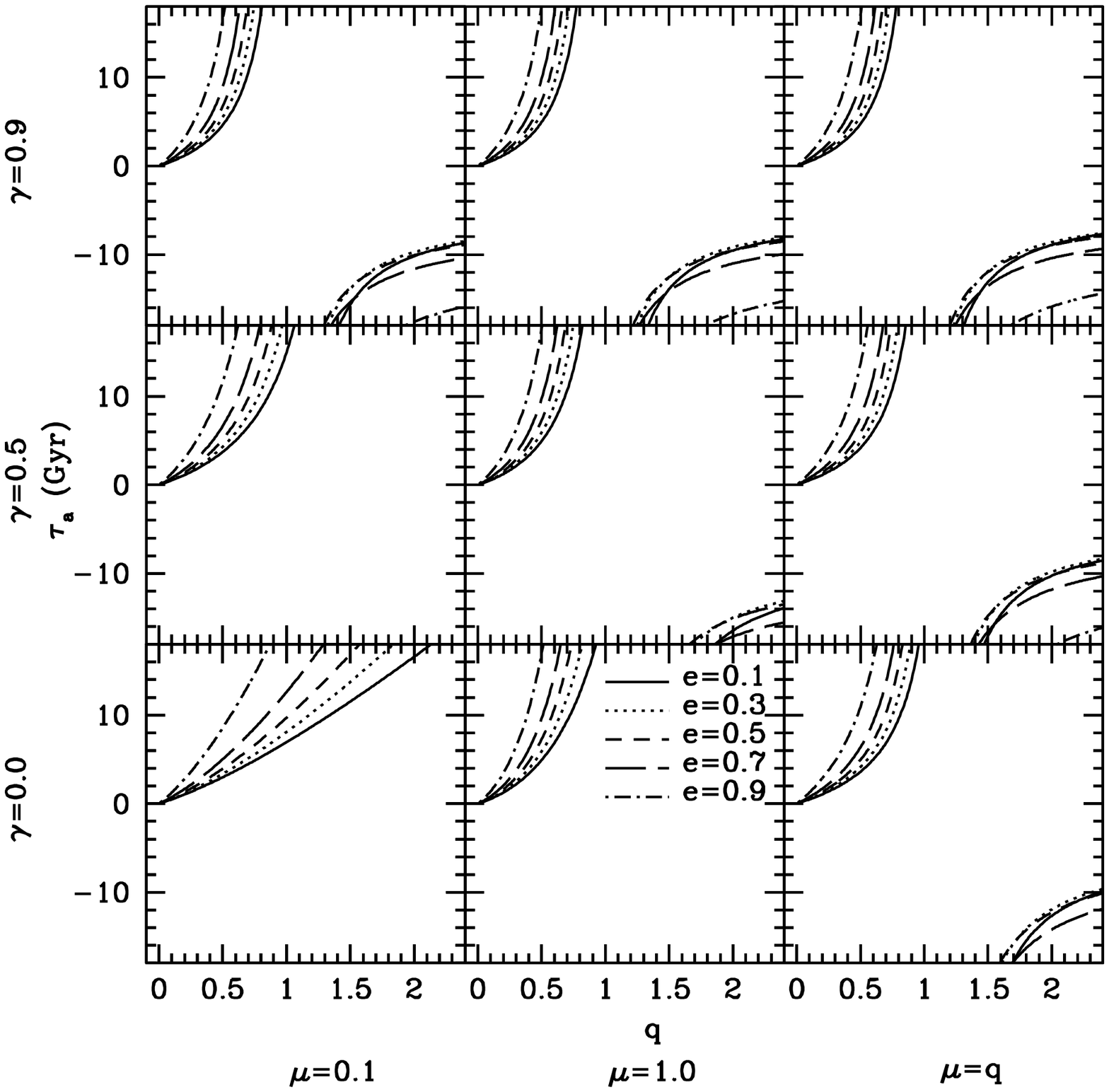}}
\resizebox{8.5cm}{!}{\includegraphics{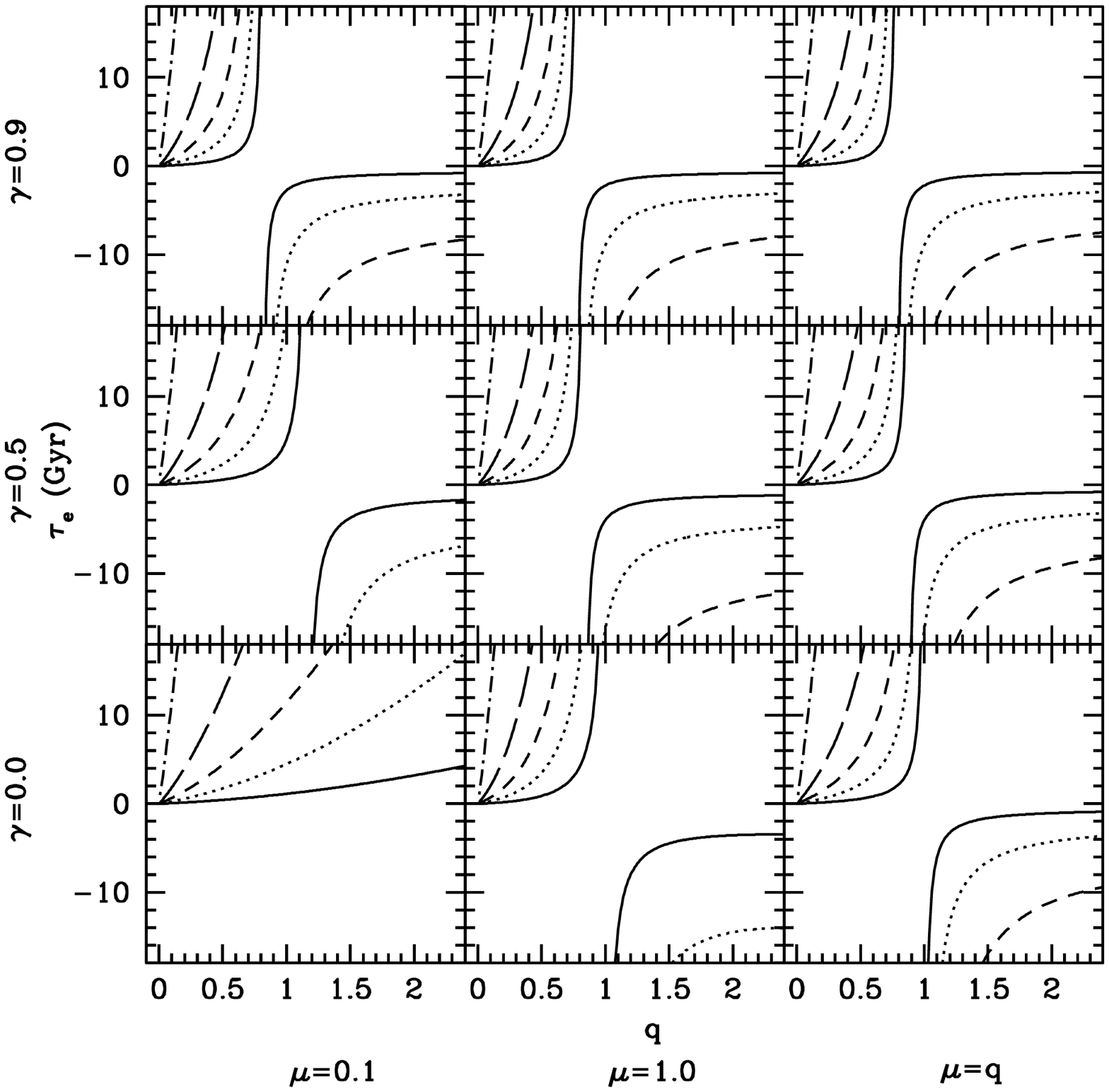}}
\end{center}
\caption{Orbital evolution timescales for the semi-major axis $a$ ({\it
top}) and orbital eccentricity $e$ ({\it bottom}) for a delta function
mass loss rate $\dot{M}_1 = \dot{M}_0\, \delta(\nu)$ with
$\dot{M}_0=-10^{-9}\,M_\odot\,{\rm yr}^{-1}$ for a range of 
eccentricities $e=0.1$, 0.3, 0.5, 0.7, 0.9, as shown in the bottom 
middle panel of the left plot. The timescales are
calculated as a function of the binary mass ratio $q=M_1/M_2$ under the
assumptions that the donor star rotates synchronously with the orbital
angular velocity at periastron and that the accretor is a neutron star
of mass $M_2=1.44\, M_\odot$. The different panels show timescales for
different values of $\gamma$ and $\mu$. The $\mu=q$ panels correspond to
the case where the matter lost from the system carries the specific
orbital angular momentum of the accretor.  Negative timescales 
correspond to a decreasing semi-major axis or eccentricity, while 
positive timescales correspond to semi-major axis or eccentricity 
growth.}
\label{fig-ta}
\end{figure*}

In Figure~\ref{fig-ta}, the evolutionary timescales 
$\tau_a = a/\dot{a}$ and $\tau_e = e/\dot{e}$ for the semi-major axis 
$a$ and orbital eccentricity $e$ due to mass transfer in an eccentric 
binary are shown as a function of the binary mass ratio $q$, for 
different values of the orbital eccentricity and the parameters $\gamma$ 
and $\mu$ [see Eqs.~(\ref{gammadef}) and~(\ref{mudef})]. While the
actual timescales are given by the absolute values of $\tau_a$
and $\tau_e$, we here allow the timescales to be negative as well as
positive to distinguish between negative and positive rates of
change of the orbital elements. 

For the calculation of the timescales, we assume the donor to rotate
synchronously with the orbital angular velocity at periastron, and take
the accretor to be a $1.44\,M_\odot$ neutron star. The donor mass is
then fixed by the binary mass ratio. Since the radius of the neutron
star is much smaller than the semi-major axis of the orbit, the terms in
Eqs.~(\ref{eq-deltaa}) and (\ref{eq-deltae}) containing the ratio
$|\vec{r}_{A_2}|/a$ are negligible compared to the other terms. We
therefore set $|\vec{r}_{A_2}|=0$, so that the timescales $\tau_a$ and
$\tau_e$ are independent of $a$. For the mass transfer rate we adopt a
constant $\dot{M}_0 = -10^{-9}\, M_\odot\, {\rm yr^{-1}}$, but note that
the linear dependence of the rates of change of the orbital semi-major
axis and eccentricity on $\dot{M}_0$ allows for an easy rescaling of the
timescales to different mass transfer rates. 

The overall shape of the curves shown in Figure~\ref{fig-ta} is similar
to that of the curves shown in Figure~2 of Paper~I in the case of
conservative mass transfer. For a given fraction $\gamma$ of mass loss
from the system, increasing $\mu$ implies more angular momentum loss
from the binary, and thus faster orbital shrinkage or slower orbital
expansion.  For a given degree $\mu$ of specific orbital angular
momentum loss, on the other hand, decreasing $\gamma$ implies more mass
loss from the binary, causing faster orbital expansion and slower
orbital contraction. The timescales for the evolution of the orbital
eccentricity show a similar dependence on the parameters $\gamma$ and
$\mu$. 

\begin{figure*}
\begin{center}
\resizebox{8.5cm}{!}{\includegraphics{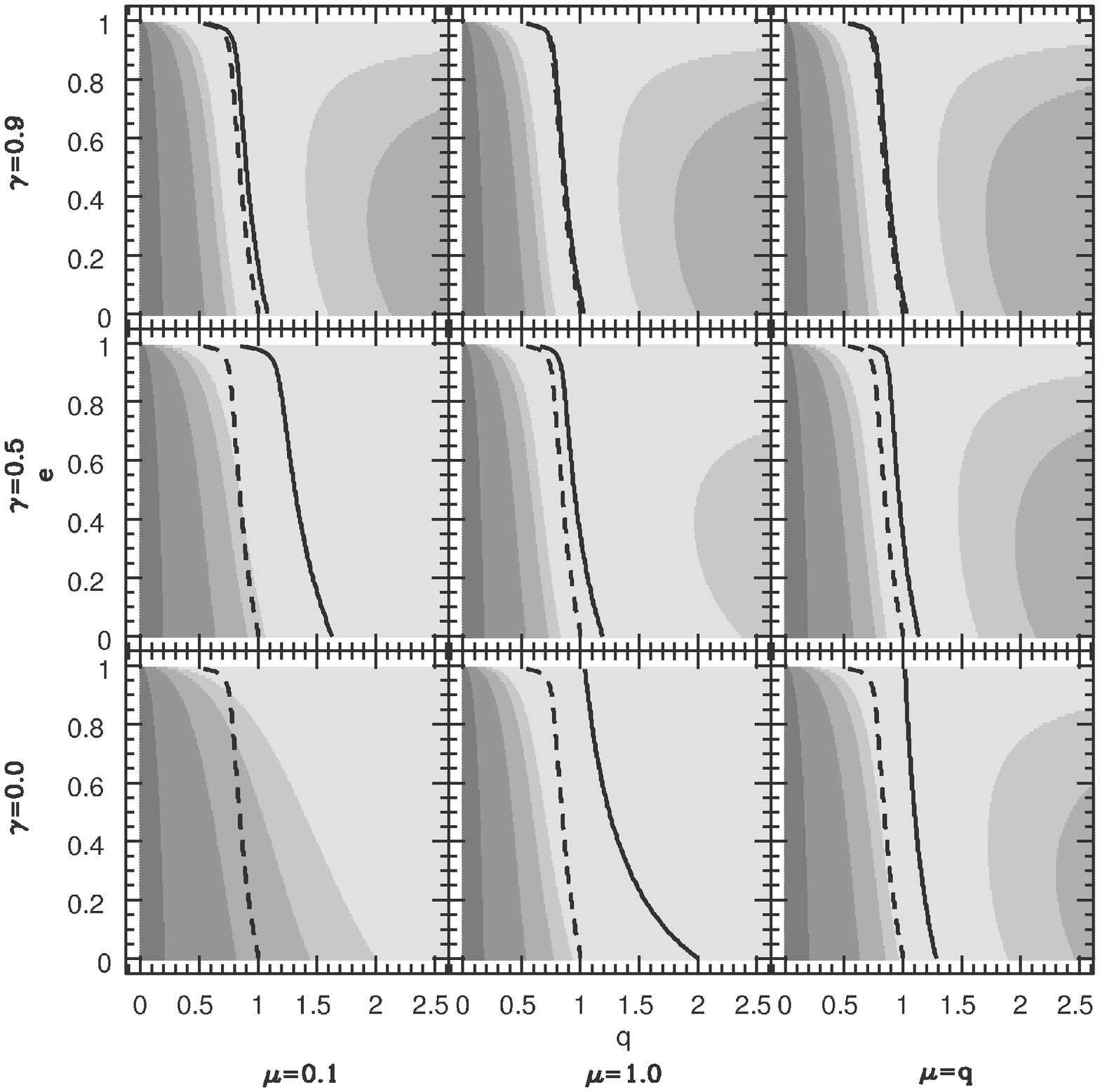}}
\resizebox{8.5cm}{!}{\includegraphics{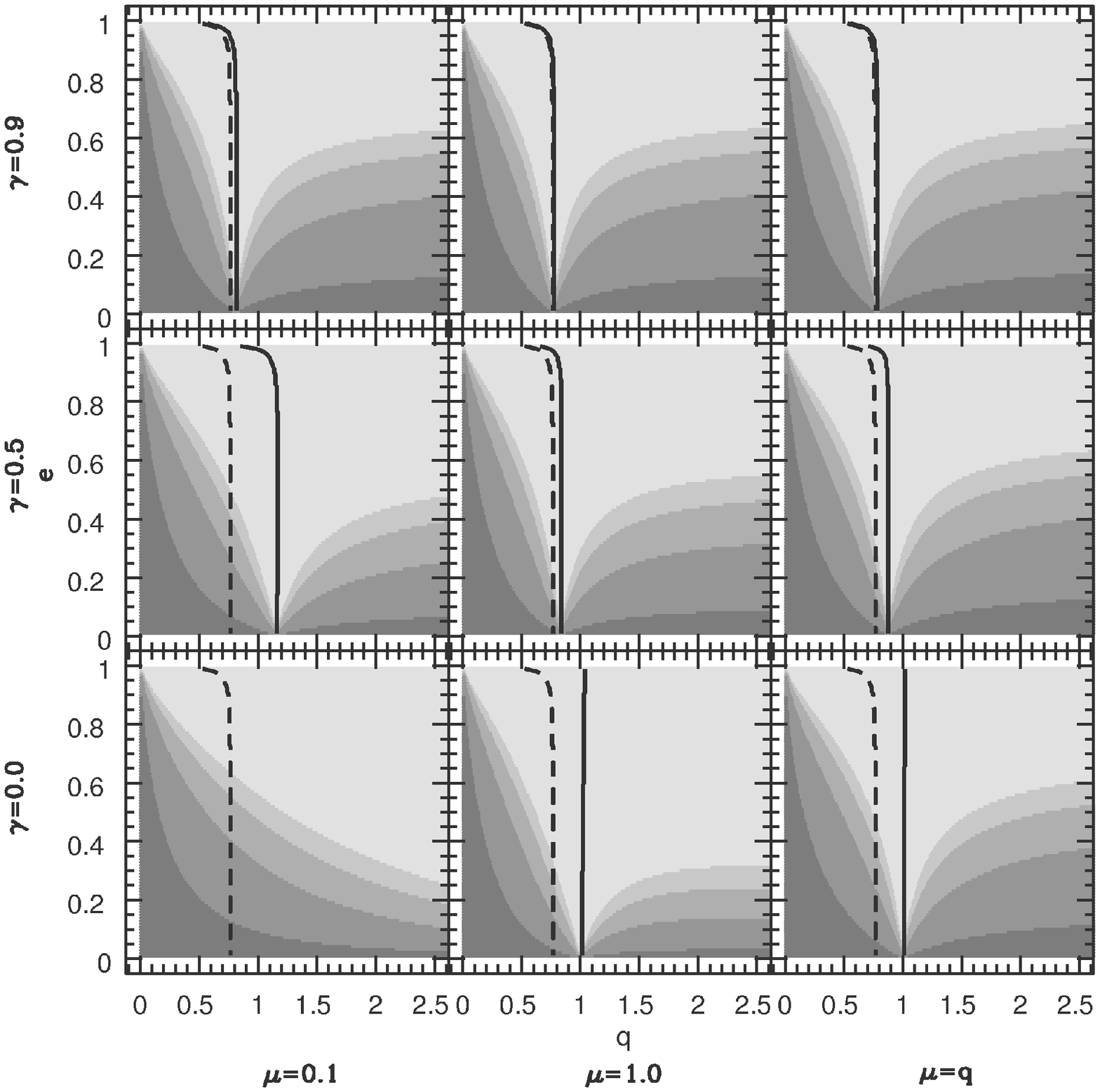}}
\end{center}
\caption{Contour plots of the orbital evolution timescales for the
semi-major axis $a$ ({\it top}) and orbital eccentricity $e$ ({\it
bottom}) in the $(q,e)$-plane for the same set of assumptions as adopted
in Fig.~\ref{fig-ta}. From the darkest to the lightest gray, the
different gray shades represent timescales from 0 to 1\,Gyr, 1 to
5\,Gyr, 5 to 10\,Gyr, 10 to 15\,Gyr, and more than 15\,Gyr,
respectively. The thick black line in each panel separates the regions
of the $(q,e)$ space where $\dot{a} > 0$ or $\dot{e} > 0$ ({\it left of
the thick black line}) from the regions of the $(q,e)$ space where
$\dot{a} < 0$ or $\dot{e} < 0$ ({\it right of the thick black line}).
The timescales in the bottom left panel ($\gamma=0$ and $\mu=0.1$) are
positive for all values of $q$ and $e$ displayed. For comparison, the
dashed black line shows the dividing line between between increasing and
decreasing orbital elements in the case of fully conservative mass
transfer.}
\label{fig-cta}
\end{figure*}

The evolutionary timescales of the orbital semi-major axis $a$ and
eccentricity $e$ can be negative as well as positive, depending on the
initial binary properties. The transition from shrinking to growing
orbital elements is illustrated more clearly by the contour plots shown
in Figure~\ref{fig-cta}. In these plots, the gray shades represent
different timescales of orbital evolution and the thick black line marks
the transition from shrinking ({\it right of the thick black line}) to
growing ({\it left of the thick black line}) orbital elements. For
comparison, the dashed black line marks the transition in the case of
fully conservative mass transfer presented in Paper~I\footnote{The sharp
bend of the dashed line toward smaller $q$ values for $e \la 0.95$ was
not observed in Paper~I due to the lower resolution of the orbital
eccentricity grid considered in that paper. Equations (41) and (42) of
Paper~I still fit the line to better than 1.5\% (10\%) for $e \la 0.8$
($e \la 0.95$) in the case of the orbital semi-major axis and for $e \la
0.85$ ($e \la 0.95$) in the case of the orbital eccentricity.}. As
$\gamma$ decreases, more mass is lost from the system and the transition
from negative to positive rates of change of the semi-major axis and
eccentricity moves to larger mass ratios. Conversely, as $\mu$
increases, more angular momentum is lost from the system and the
transition from negative to positive rates of change of the semi-major
axis and eccentricity moves to smaller mass ratios. In the particular
case where matter leaving the system carries away the specific orbital
angular momentum of the accretor ($\mu=q$), the critical mass ratio
separating increasing from decreasing orbital semi-major axes and
eccentricities increases by about 30\% when going from fully
conservative to fully non-conservative mass transfer. 

To assess the role of mass transfer in the overall evolution of the
binary, we compare the timescales shown in Figures~\ref{fig-ta}
and~\ref{fig-cta} with the orbital evolution timescales due to tidal
dissipation in the Roche-lobe filling star. The tidal evolution
timescales are determined as in \citet{HTP02} assuming the donor
is a zero-age main-sequence star. The variations of the tidal evolution
timescales as a function of the orbital eccentricity and binary mass
ratio are shown in Figure 4 of Paper~I. The discontinuity of the
timescales at $q \simeq 0.87$ corresponds to the transition from donor
stars in which convective damping is the dominant tidal dissipation
mechanism ($M_2 \la 1.25\,M_\odot$) to donor stars in which radiative
damping is the dominant tidal dissipation mechanism ($M_2 \ga
1.25\,M_\odot$). 

\begin{figure*}
\begin{center}
\resizebox{8.5cm}{!}{\includegraphics{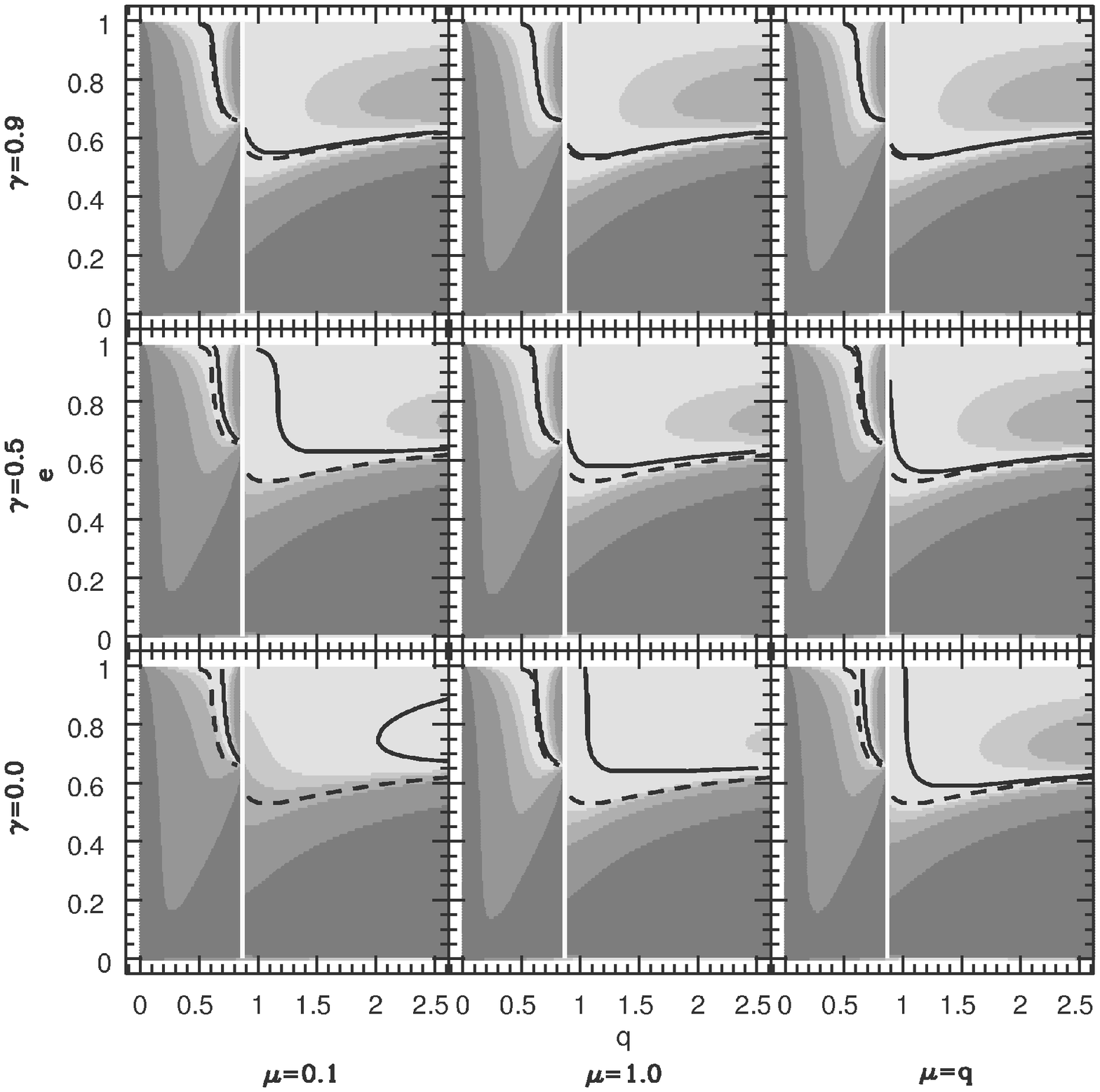}}
\resizebox{8.5cm}{!}{\includegraphics{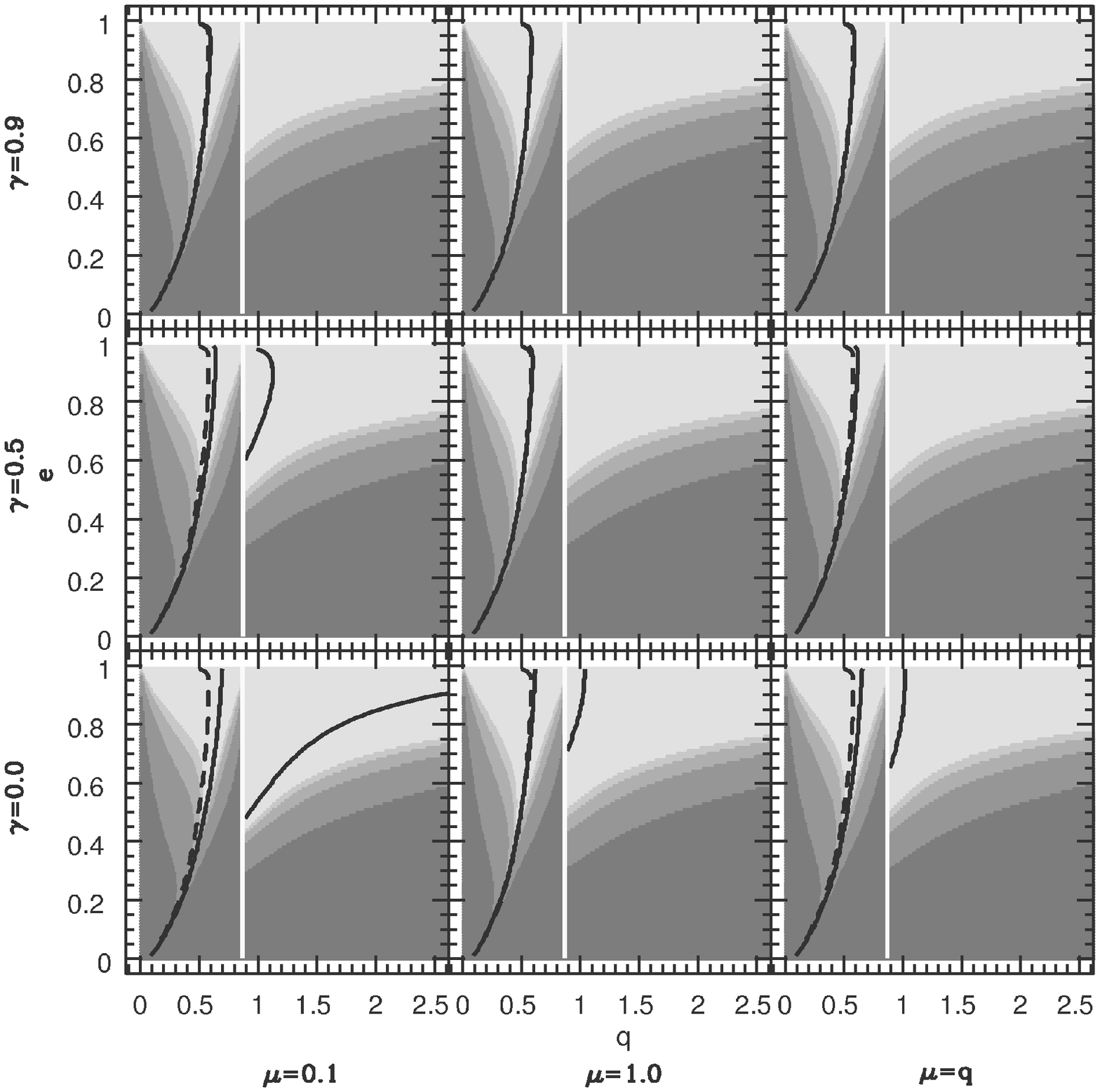}}
\end{center}
\caption{Contour plots of the orbital evolution timescales for the
semi-major axis $a$ ({\it top}) and orbital eccentricity $e$ ({\it
bottom}) in the $(q,e)$-plane due to the combined effects of mass
transfer and tidal dissipation in the donor star of a semi-detached
binary. The mass transfer rate is assumed to be a delta function of
amplitude $\dot{M}_0=-10^{-9}\,M_\odot\,{\rm yr}^{-1}$ at the periastron
of the binary orbit, the donor star is assumed to rotate synchronously
with the orbital angular velocity at periastron, and the accretor is
assumed to be a $1.44\, M_\odot$ neutron star. The tidal contribution to
the orbital evolution timescales is determined assuming the donor star
has a radius equal to that of a zero-age main-sequence star of the
considered mass. From the darkest to the lightest gray, the different
gray shades represent timescales from 0 to 1\,Gyr, 1 to 5\,Gyr, 5 to
10\,Gyr, 10 to 15\,Gyr, and more than 15\,Gyr, respectively. The thick
black line in each panel separates the regions of the $(q,e)$ space
where $\dot{a} > 0$ or $\dot{e} > 0$ ({\it left of the thick black
line}) from the regions of the $(q,e)$ space where $\dot{a} < 0$ or
$\dot{e} < 0$ ({\it right of the thick black line}). For comparison, the
dashed black line shows the dividing line between between increasing and
decreasing orbital elements in the case of fully conservative mass
transfer. The vertical white line at $q \approx 0.87$ indicates the
transition from convective damping (for $M_2 \la 1.25\,M_\odot$) to
radiative damping (for $M_2 \ga 1.25\,M_\odot$) as the dominant energy
dissipation mechanism.}
\label{fig-ctaT}
\end{figure*}

The timescales of orbital evolution due to the combined effects of mass
transfer and tides are shown as contour plots in Figure~\ref{fig-ctaT}.
As before, the thick black line indicates the transition from shrinking
({\it right of the thick black line}) to growing ({\it left of the thick
black line}) orbital semi-major axis and eccentricity. For comparison,
the transition line for conservative mass transfer is shown by means of
the dashed black line. The vertical white line at $q \simeq 0.87$
separates donor stars in which tidal energy is dissipated by convective
damping from donor stars in which tidal energy is dissipated by
radiative damping. As noted in Paper~I, in the case of conservative mass
transfer, there are large regions in the $(q,e)$ parameter space where
the combined effects of mass transfer and tidal interactions do {\it
not} lead to rapid circularization of the orbit after the onset of mass
transfer and where eccentricity pumping occurs instead of eccentricity
damping. When mass loss from the system is taken into account, the
parameter space for eccentricity pumping becomes even larger, though the
timescales of orbital evolution in the newly accessible $\dot{e} > 0$
regions are long ($\ga 15$\,Gyr). Similar behavior is observed for the
evolution of the orbital semi-major axis. 

We note that the results for zero-age main-sequence stars presented here 
are intended for illustration and comparison with Paper I. As tides will 
be much more efficiently damped in giant stars with deep convective 
envelopes, the interplay between mass transfer and tidal interactions 
will likely change significantly with the evolutionary state of the 
donor star.


\section{Concluding Remarks}

We extended the formalism to study the orbital evolution due to mass
transfer in eccentric binaries derived in Paper~I to account for the
effects of mass and angular momentum loss from the system. Adopting a
delta function mass transfer rate at the periastron of the binary orbit,
we find that the usually adopted assumption of rapid orbital
circularization during the early stages of mass transfer remains
unjustified when systemic mass and angular momentum loss are taken into
account. Our results thus present a possible explanation for the 
observation of non-zero orbital eccentricities in mass-transferring 
binaries.

The formalism presented in this paper and in Paper~I can be incorporated
into binary evolution and population synthesis code to provide a model
for eccentric mass-transferring binaries which are currently, by
construction, absent in any population synthesis studies of interacting
binaries and their descendants. In future work, we intend to 
incorporate this orbital evolution into such code, and determine the 
prevalence and the long-term evolutionary effects of mass transfer 
in eccentric binary systems.  Possible applications include the
modeling of systems such as Cir\,X-1 in which a neutron star is thought
to accrete matter from a Roche-lobe filling or nearly Roche-lobe filling
companion during each periastron passage and in which near-IR and X-ray
spectroscopy support the presence of an accretion-driven mass outflow
\citep{2003A&A...400..655C, 2008ApJ...673.1033I, 1986MNRAS.221P..27T}.
Another example is the ultracompact X-ray binary 4U\,1820-30 which is
thought to be a member of a hierarchical triple
\citep{2001ApJ...563..934C, 2007MNRAS.377.1006Z}.  Furthermore, dynamical
interactions between single and binary stars in dense stellar clusters
can induce eccentricities in circular binaries and enhance
eccentricities in already eccentric binaries
\citep{1996MNRAS.282.1064H}.  This induced eccentricity can directly
lead to Roche Lobe overflow at periastron, as has been suggested for the
flaring X-ray binaries in NGC~4697 \citep{2005MNRAS.364..971M}.  In
future work, we intend to model the mass transfer rate at periastron
more realistically by taking into account the atmospheric properties of
the donor star and considering the feedback of the orbital and radial
evolution of the star on the mass transfer rate.

\acknowledgements

This work is partially supported by NSF Award AST-0525995//ASW01
(subcontract from Adler Planetarium and Astronomy Museum), NSF CAREER
Award AST-0449558, and NASA BEFS Award NNG06GH87G to VK.  F.A.R. 
acknowledges support from NASA Grant NNG06GI62G.  Numerical simulations
presented in this paper were performed on the HPC cluster {\it fugu}
available to the Theoretical Astrophysics Group at Northwestern
University through NSF MRI grant PHY-0619274 to VK.


\bibliography{MT_noncons}

\begin{thebibliography}{19}
\expandafter\ifx\csname natexlab\endcsname\relax\def\natexlab#1{#1}\fi

\bibitem[{{Brouwer} \& {Clemence}(1961)}]{BC61}
{Brouwer}, D., \& {Clemence}, G.~M. 1961, {Methods of celestial mechanics} (New
  York: Academic Press, 1961)

\bibitem[{{Chou} \& {Grindlay}(2001)}]{2001ApJ...563..934C}
{Chou}, Y., \& {Grindlay}, J.~E. 2001, \apj, 563, 934

\bibitem[{{Clark} {et~al.}(2003){Clark}, {Charles}, {Clarkson}, \&
  {Coe}}]{2003A&A...400..655C}
{Clark}, J.~S., {Charles}, P.~A., {Clarkson}, W.~I., \& {Coe}, M.~J. 2003,
  \aap, 400, 655

\bibitem[{{Danby}(1962)}]{D62}
{Danby}, J. 1962, {Fundamentals of celestial mechanics} (New York: Macmillan,
  1962)

\bibitem[{{Fitzpatrick}(1970)}]{F70}
{Fitzpatrick}, P.~M. 1970, {Principles of celestial mechanics} (New York,
  Academic Press [1970])

\bibitem[{{Hadjidemetriou}(1969)}]{1969Ap&SS...3..330H}
{Hadjidemetriou}, J.~D. 1969, \apss, 3, 330

\bibitem[{{Heggie} \& {Rasio}(1996)}]{1996MNRAS.282.1064H}
{Heggie}, D.~C., \& {Rasio}, F.~A. 1996, \mnras, 282, 1064

\bibitem[{{Hurley} {et~al.}(2002){Hurley}, {Tout}, \& {Pols}}]{HTP02}
{Hurley}, J.~R., {Tout}, C.~A., \& {Pols}, O.~R. 2002, \mnras, 329, 897

\bibitem[{{Iaria} {et~al.}(2008){Iaria}, {D'A{\'{\i}}}, {Lavagetto}, {Di
  Salvo}, {Robba}, \& {Burderi}}]{2008ApJ...673.1033I}
{Iaria}, R., {D'A{\'{\i}}}, A., {Lavagetto}, G., {Di Salvo}, T., {Robba},
  N.~R., \& {Burderi}, L. 2008, \apj, 673, 1033

\bibitem[{{Kolb} {et~al.}(2001){Kolb}, {Rappaport}, {Schenker}, \&
  {Howell}}]{Kolb2001}
{Kolb}, U., {Rappaport}, S., {Schenker}, K., \& {Howell}, S. 2001, \apj, 563,
  958

\bibitem[{{Maccarone}(2005)}]{2005MNRAS.364..971M}
{Maccarone}, T.~J. 2005, \mnras, 364, 971

\bibitem[{{Matese} \& {Whitmire}(1983)}]{MW83}
{Matese}, J.~J., \& {Whitmire}, D.~P. 1983, \apj, 6, 776

\bibitem[{{Petrova} \& {Orlov}(1999)}]{1999AJ....117..587P}
{Petrova}, A.~V., \& {Orlov}, V.~V. 1999, \aj, 117, 587

\bibitem[{{Raguzova} \& {Popov}(2005)}]{2005A&AT...24..151R}
{Raguzova}, N.~V., \& {Popov}, S.~B. 2005, Astronomical and Astrophysical
  Transactions, 24, 151

\bibitem[{{Sepinsky} {et~al.}(2007{\natexlab{a}}){Sepinsky}, {Willems}, \&
  {Kalogera}}]{2007ApJ...660.1624S}
{Sepinsky}, J.~F., {Willems}, B., \& {Kalogera}, V. 2007{\natexlab{a}}, \apj,
  660, 1624

\bibitem[{{Sepinsky} {et~al.}(2007{\natexlab{b}}){Sepinsky}, {Willems},
  {Kalogera}, \& {Rasio}}]{2007ApJ...667.1170S}
{Sepinsky}, J.~F., {Willems}, B., {Kalogera}, V., \& {Rasio}, F.~A.
  2007{\natexlab{b}}, \apj, 667, 1170

\bibitem[{{Sterne}(1960)}]{S60}
{Sterne}, T.~E. 1960, {An introduction to celestial mechanics} (Interscience
  Tracts on Physics and Astronomy, New York: Interscience Publication, 1960)

\bibitem[{{Tennant} {et~al.}(1986){Tennant}, {Fabian}, \&
  {Shafer}}]{1986MNRAS.221P..27T}
{Tennant}, A.~F., {Fabian}, A.~C., \& {Shafer}, R.~A. 1986, \mnras, 221, 27P

\bibitem[{{Zdziarski} {et~al.}(2007){Zdziarski}, {Wen}, \&
  {Gierli{\'n}ski}}]{2007MNRAS.377.1006Z}
{Zdziarski}, A.~A., {Wen}, L., \& {Gierli{\'n}ski}, M. 2007, \mnras, 377, 1006

\end{thebibliography}

\end{document}